\documentclass[journal]{IEEEtran}
\pdfoutput=1

\usepackage[utf8]{inputenc}
\usepackage[T1]{fontenc}
\usepackage{microtype}

\usepackage{amsmath}
\usepackage{amsfonts}

\usepackage{graphicx}
\graphicspath{{figs/}}
 
\usepackage{booktabs}
\usepackage{multirow}

\usepackage{algorithm}
\usepackage[noEnd=false,indLines=true]{algpseudocodex}

\allowdisplaybreaks[2]

\usepackage{hyperref}
\usepackage{url}

\usepackage{etoolbox}
\makeatletter
\expandafter\patchcmd\csname\string\IEEEbiography\endcsname
  {\vskip \@IEEEBIOskipN plus 1fil minus 0\baselineskip}
  {\vskip \@IEEEBIOskipN plus 1\baselineskip minus 0\baselineskip}
  {\typeout{IEEEbiography skip patch applied}}
  {\ClassWarning{IEEEtran}{IEEEbiography skip patch FAILED}}
\makeatother

\begin{document}

\title{Accelerating Underground Pumped Hydro Energy Storage Scheduling with Decision-Focused Learning}

\author{Honghui~Zheng,~\IEEEmembership{Student Member,~IEEE,}
        Pietro~Favaro,
        Yury~Dvorkin,~\IEEEmembership{Member,~IEEE,}
        and~Ján~Drgoňa,~\IEEEmembership{Member,~IEEE}
\thanks{H. Zheng, Y. Dvorkin, and J. Drgoňa are with the Department of Civil and Systems Engineering, Johns Hopkins University, Baltimore, MD 21218, USA (e-mail: \{hzheng39, ydvorki1, jdrgona1\}@jh.edu).}
\thanks{P. Favaro is with the Power Systems and Markets Research Group, University of Mons, 7000 Mons, Belgium (e-mail: pietro.favaro@umons.ac.be).}
\thanks{This article has been accepted for publication in IEEE Transactions on Sustainable Energy. \textcopyright~2026 IEEE. Personal use of this material is permitted. Permission from IEEE must be obtained for all other uses, in any current or future media, including reprinting/republishing this material for advertising or promotional purposes, creating new collective works, for resale or redistribution to servers or lists, or reuse of any copyrighted component of this work in other works. DOI: \href{https://doi.org/10.1109/TSTE.2026.3722492}{10.1109/TSTE.2026.3722492}. \textit{(Corresponding author: Honghui Zheng.)}}}

\markboth{IEEE Transactions on Sustainable Energy}%
{Zheng \MakeLowercase{\textit{et al.}}: Accelerating Underground Pumped Hydro Energy Storage Scheduling with Decision-Focused Learning}

\maketitle

\begin{abstract}
Underground pumped hydro energy storage (UPHES) systems play a critical role in grid-scale energy storage for renewable integration, yet optimal day-ahead scheduling remains computationally prohibitive due to nonlinear turbine performance characteristics and discrete operational modes. This paper presents a decision-focused learning (DFL) framework that addresses the computational-accuracy trade-off in UPHES day-ahead scheduling. The proposed methodology employs neural networks to predict penalty weights that guide recursive linearization, transforming the intractable MINLP into a sequence of convex quadratic programs trained end-to-end via differentiable optimization layers. Case studies across 19 representative Belgian electricity market scenarios demonstrate that the DFL framework effectively navigates the trade-off between solution quality and computation time. As a refiner, DFL adds $\approx$1.2~s of post-processing to improve profit by 1.1\% over piecewise MIQP baselines. As a real-time scheduler initialized with a fast global linear MIQP, the end-to-end time is 3.87~s (a 300-fold speedup) while maintaining profitability within 3.6\% of the piecewise MIQP benchmark. An MIP-free deployment mode, which warm-starts from a historical schedule database and thus needs no mixed-integer solve, retains 92\% of benchmark profit on held-out price scenarios.
\end{abstract}


\section*{Nomenclature}
\addcontentsline{toc}{section}{Nomenclature}

\subsection*{Sets and Indices}
\begin{IEEEdescription}[\IEEEusemathlabelsep\IEEEsetlabelwidth{$\mathcal{M}_{\mathrm{act}}$}]
\item[$\mathcal{T}$] Set of time periods in the planning horizon.
\item[$t$] Time index representing hourly intervals, $t \in \mathcal{T}$.
\item[$\mathcal{M}$] Set of operating modes, $\mathcal{M} = \{I, T, P\}$ (idle, turbine, pump).
\item[$m$] Operating-mode index, $m \in \mathcal{M}$.
\item[$\mathcal{I}_h$] Set of head discretization points for SOS2.
\item[$\mathcal{I}_p^m$] Set of power discretization points in SOS2 for $m$.
\item[$i, j$] Indices for discretization points in SOS2.
\end{IEEEdescription}

\subsection*{Decision Variables}
\begin{IEEEdescription}[\IEEEusemathlabelsep\IEEEsetlabelwidth{$\Omega_{t,i,j}^m$}]
\item[$p_t^m$] Power output (turbine) or consumption (pump) at time $t$ for mode $m \in \{T, P\}$ [MW].
\item[$q_t^m$] Water flow rate at time $t$ for mode $m$ [m$^3$/s].
\item[$h_t$] Hydraulic head at time $t$ [m].
\item[$v_{\text{low},t}$] Lower reservoir volume at time $t$ [m$^3$].
\item[$z_t^m$] Binary status variable for mode $m \in \mathcal{M}$ at time $t$.
\item[$\Omega_{t,i}^{\text{vh}}$] SOS2 interpolation weight for volume-head relationship at time $t$, discretization point $i$.
\item[$\Omega_{t,i,j}^m$] SOS2 interpolation weight for UPC of mode $m$ at time $t$, grid point $(i,j)$.
\end{IEEEdescription}

\subsection*{System Parameters}
\begin{IEEEdescription}[\IEEEusemathlabelsep\IEEEsetlabelwidth{$h_\text{min}, h_\text{max}$}]
\item[$\lambda_t^{\text{DA}}$] Day-ahead electricity price at time $t$ [€/MWh].
\item[$C_{\text{op}}$] Operational cost coefficient [€/MW$^2$].
\item[$\Delta t$] Time step duration, 3600 s.
\item[$v_{\text{low}}^{\text{init}}$] Initial lower reservoir volume [m$^3$].
\item[$v_{\text{low}}^{\text{target}}$] Target lower reservoir volume [m$^3$].
\item[$v_{\text{total}}$] Total combined capacity of reservoirs [m$^3$].
\item[$h_\text{min}, h_\text{max}$] Minimum and maximum hydraulic head [m].
\item[$h^{\text{init}}$] Initial hydraulic head [m].
\end{IEEEdescription}

\subsection*{Nonlinear Functions and Mappings}
\begin{IEEEdescription}[\IEEEusemathlabelsep\IEEEsetlabelwidth{$f_m^{\text{UPC}}(p, h)$}]
\item[$f_m^{\text{UPC}}(p, h)$] Unit performance curve for mode $m$ [m$^3$/s].
\item[$p_{\min}^m(h)$] Minimum power for mode $m$ [MW].
\item[$p_{\max}^m(h)$] Maximum power for mode $m$ [MW].
\item[$f_{\text{up}}^{\text{vol}}(h)$] Upper reservoir volume function [m$^3$].
\item[$f_{\text{low}}^{\text{vol}}(h)$] Lower reservoir volume function [m$^3$].
\item[$c_{ab}^{(m)}$] Polynomial coefficients for UPC of mode $m$.
\item[$d$] Total polynomial degree for UPC.
\end{IEEEdescription}

\subsection*{Linearization Coefficients}
\begin{IEEEdescription}[\IEEEusemathlabelsep\IEEEsetlabelwidth{$\boldsymbol{\beta}_m^{\max}$}]
\item[$\boldsymbol{\alpha}_m$] Coefficient vector for global linear approximation of UPC for mode $m$ [m$^3$/s per (MW, m)].
\item[$\boldsymbol{\beta}_m^{\min}$] Coefficient vector for linearization of minimum power boundary, mode $m$ [MW per m].
\item[$\boldsymbol{\beta}_m^{\max}$] Coefficient vector for linearization of maximum power boundary, mode $m$ [MW per m].
\item[$\boldsymbol{\delta}$] Coefficient vector for global linear approximation of volume-head relationship [m per m$^3$].
\item[$\boldsymbol{\xi}_{q,t}^{(k)}$] Local linearization coefficient vector for UPC at time $t$, iteration $k$.
\item[$\boldsymbol{\xi}_{v,t}^{(k)}$] Local linearization coefficient vector for volume-head at time $t$, iteration $k$.
\end{IEEEdescription}

\subsection*{Sample Points (Piecewise Interpolation)}
\begin{IEEEdescription}[\IEEEusemathlabelsep\IEEEsetlabelwidth{$q_{i,j}^{m,\text{sample}}$}]
\item[$h_i^{\text{sample}}$] Sampled head value at point $i$ [m].
\item[$v_{\text{low},i}^{\text{sample}}$] Sampled lower reservoir volume at point $i$ [m$^3$].
\item[$p_{i,j}^{m,\text{sample}}$] Sampled power value at point $(i,j)$ for $m$ [MW].
\item[$q_{i,j}^{m,\text{sample}}$] Sampled flow rate at point $(i,j)$ for $m$ [m$^3$/s].
\end{IEEEdescription}

\subsection*{Decision-Focused Learning Parameters}
\begin{IEEEdescription}[\IEEEusemathlabelsep\IEEEsetlabelwidth{$\mathbf{w}_p, \mathbf{w}_q, \mathbf{w}_h$}]
\item[$\theta$] Neural network parameters.
\item[$\mathbf{w}_p, \mathbf{w}_q, \mathbf{w}_h$] Penalty weights for power, flow, and head.
\item[$K$] Total number of recursive linearizations.
\item[$\gamma > 1$] Penalty growth factor.
\end{IEEEdescription}

\subsection*{Performance Metrics}
\begin{IEEEdescription}[\IEEEusemathlabelsep\IEEEsetlabelwidth{$\text{Vol}$}]
\item[$\Pi$] Ex-post profit evaluated using simulated trajectories under true nonlinear dynamics [EUR].
\item[$\text{SI}$] System imbalance penalty for power schedule [EUR].
\item[$\text{Vol}$] Penalty for exceeding target volume [EUR].
\item[$\mathcal{L}(\theta)$] Training loss function (negative ex-post profit).
\end{IEEEdescription}

\subsection*{Solution Notation Convention}
\begin{IEEEdescription}[\IEEEusemathlabelsep\IEEEsetlabelwidth{$\hat{\mathbf{x}}^{(k)}$}]
\item[$\bar{\mathbf{x}}$] Initial noisy solution from baseline MIQP.
\item[$\hat{\mathbf{x}}^{(k)}$] Optimized solution at iteration $k$.
\item[$\tilde{\mathbf{x}}$] Simulated solution under true nonlinear dynamics.
\end{IEEEdescription}

\section{Introduction} 
\label{sec:introduction}

\IEEEPARstart{G}{rid}-scale energy storage is critical to compensate for the temporal variance of weather-dependent generation, such as wind, solar, and loads~\cite{golshani2018coordination}. Despite the recent proliferation of battery storage systems~\cite{wang2016review}, pumped hydro energy storage (PHES) is anticipated to remain the dominant form of grid-scale energy storage globally, particularly for durations of four hours and more. As a proven technology with decades of operational history~\cite{REHMAN2015586}, PHES accounts for the vast majority of installed large-scale storage capacity, representing about 99\% of the world’s electricity storage capacity and around 3\% of the total installed electricity generation capacity globally~\cite{rastler2010electricity}, far surpassing -- at least for the present moment -- the contribution of battery energy storage systems~\cite{blakers2021review}. Both the operational maturity and deployment scale have established PHES as the backbone of grid-scale energy storage, providing reliable bulk energy shifting and ancillary services~\cite{hunt2020existing}.

Although conventional PHES installations require specific landscapes with significant elevation differences between upper and lower reservoirs, Underground Pumped Hydro Energy Storage (UPHES) is suited for regions with limited topographical relief or environmental constraints that preclude natural surface reservoirs or their low-impact development~\cite{menendez2019pumped}. For example, by repurposing abandoned mines~\cite{winde2017exploring}, and aquifers~\cite{martin2007aquifer} as lower reservoirs, UPHES circumvents such geographical limitations and reduces environmental impacts and land-use requirements~\cite{yang2016pumped}. UPHES also has the potential to transform decommissioned mining facilities into storage assets~\cite{pickard2011history}, which in turn may support local communities~\cite{TASSI2024114467, COLAS2023109153}.

Operating the UPHES system efficiently is a substantial computational challenge due to nonlinear physical dynamics and relatively large round-trip inefficiencies~\cite{emrani2022assessment}, discrete operational turbine and pumping modes~\cite{jia2013mixed}, and optimization under uncertain electricity market conditions~\cite{benini2002day, 7353197}. The original computational challenge stems from the Unit performance curves (UPCs) which relate hydraulic head (elevation difference between upper and lower reservoirs), power output, and water flow via a complex nonlinear and non-convex mapping~\cite{finardi2016assessing}. Furthermore, underground reservoir geometries introduce discontinuous head variations as water levels transition between horizontal galleries, creating stepwise changes in available head and turbine efficiency~\cite{toubeau2019non}. The system must transition among discrete operational states (pumping, generating, and idle) subject to mechanical constraints and market incentives that vary hourly or sub-hourly in electricity markets~\cite{cheng2022hierarchical}.

Existing optimization approaches for PHES scheduling predominantly employ mixed-integer programming (MIP) formulations~\cite{zhang2025long, jia2013mixed} that approximate nonlinear relationships through piecewise linearization~\cite{toubeau2019non, toubeau2019chance}. While these formulations leverage powerful solvers and provide optimality guarantees, they face fundamental trade-offs between approximation fidelity and computational tractability~\cite{andrade2022integer}. Piecewise approximations with fine discretization lead to improved solutions at the expense of substantial computational burdens that may hinder real-time application~\cite{vielma2010mixed}. Recent advances in neural network (NN)-constrained optimization~\cite{favaro2024neural} have demonstrated that trained NNs can accurately model nonconvex UPCs, but introduced additional integer variables proportional to the number of neurons in the network architecture. Despite significant progress in MIP modeling, these approximations struggle to capture nonlinearities, often yielding infeasible schedules. Alternative approaches using nonlinear optimization coupled with high-fidelity physical simulators~\cite{favaro_multi-fidelity_2024} can discover superior solutions by directly optimizing ex-post operational performance, but require substantial computational resources that preclude real-time deployment. This highlights the need for optimization frameworks that leverage learned problem structure to maintain both solution accuracy and computational efficiency. Moreover, in practice, storage scheduling is not a single one-shot routine: operators re-optimize in response to intraday forecast updates, evaluate scenario ensembles for risk management, and coordinate portfolio-level dispatch across sequential markets~\cite{KRAFT2023400}. These workflows have tight time budgets that preclude repeated MIQP solves.

Decision-focused learning (DFL) offers an approach to this challenge by directly training models to minimize downstream decision costs rather than nominal prediction errors~\cite{mandi2024decision, elmachtoub2022smart}. Donti et al.~\cite{donti2017task} showed that end-to-end task-based learning outperforms prediction-focused training on QP-based grid scheduling and energy storage arbitrage tasks, albeit under a simplified linear storage model. Traditional two-stage approaches train predictive models to minimize statistical errors under the assumption that estimation accuracy translates to optimal decisions~\cite{bertsimas2018predictive}. However, this assumption fails in complex energy systems where asymmetric cost structures and nonlinearities create non-monotonic relationships between prediction accuracy and decision quality~\cite{zhang2025decision}. Power system case studies~\cite{zhang2025decision} show certain estimation errors may incur negligible operational penalties while others critically compromise solution feasibility, yet standard two-stage models treat estimation quality uniformly~\cite{demirovic2019investigation}.

DFL addresses this limitation by embedding the optimization model within the training pipeline through differentiable optimization layers~\cite{cvxpylayers2019}, enabling end-to-end learning with loss functions derived from operational objectives rather than statistical fit. Recent power system applications demonstrate DFL's operational advantages, including wind power forecasting integrated with energy scheduling~\cite{wahdany2023more}, electricity price forecasting~\cite{tschora2023forecasting}, demand response~\cite{favaro_decision-focused_2025}, decision-focused retraining for economic dispatch~\cite{beichter2024decision, beichter2025decision}, and dispatch under decision-dependent demand uncertainty~\cite{ellinas2024decision}. Benchmark comparisons consistently show that DFL substantially reduces operational costs and decision regret relative to sequential forecasting and decision approaches~\cite{mandi2024decision, zhang2025decision}.

More recently, DFL has been applied to energy storage scheduling. Yi et al.~\cite{yi2025perturbed} embed a linear storage model as a perturbed differentiable layer for arbitrage, while Paredes et al.~\cite{paredes2025participation} formulate battery participation in reserve markets via KKT-based differentiation. Further extensions include automated feature engineering for battery scheduling~\cite{alkhulaifi2025decision} and a predict-then-bid framework with nested market-clearing layers~\cite{yi2025decision}. However, these works model storage with linear charge/discharge dynamics, whereas UPHES scheduling involves nonlinear polynomial UPCs, cubic volume--head relationships, and discrete operational modes, none of which preserve convexity. This paper accounts for these performance-critical phenomena through recursive linearization with learned trust-region weights, implementing day-ahead UPHES scheduling via an end-to-end DFL approach.

\noindent The contributions of this work are as follows:

\begin{enumerate}
    \item An end-to-end DFL framework is proposed for optimal day-ahead UPHES scheduling, utilizing differentiable optimization. A recursive local linearization technique refines an initial scheduling solution obtained from a conventional mixed-integer optimization method or a historical operation schedule.
    \item A neural penalty weight predictor is introduced to ensure solution feasibility by mapping optimization parameters to the trust region size of a given local linearization.
    \item A differentiable physics simulator is developed to directly evaluate the ex-post operational profit as training loss, ensuring solution quality over training.
    \item Case studies demonstrate that the DFL framework effectively navigates the accuracy-computation trade-off: as a refiner, it improves profit by 1.1\%; as a real-time scheduler, it achieves a 300-fold speedup while maintaining profitability within 3.6\% of the piecewise MIQP baseline; and in an MIP-free deployment mode warm-started from historical schedules (no mixed-integer solve), it retains 92\% of benchmark profit on held-out price scenarios.
    \item We open-source our code implementation of the proposed DFL framework to promote reproducibility and adoption\footnote{\url{https://github.com/SOLARIS-JHU/DFL-UPHES}}.
\end{enumerate}

\section{Problem Formulation}
\label{sec:problem}
This section presents the mathematical formulation of the UPHES scheduling problem in \eqref{eq:uphes_scheduling_problem} in the context of the day-ahead electricity market. This optimization is formulated as a Mixed-Integer Nonlinear Programming (MINLP) problem to account for nonlinear and discrete characteristics of the UPC and volume-head relationships in the lower reservoir.

\begin{figure}[t!]
    \centering
    \includegraphics[width=\columnwidth]{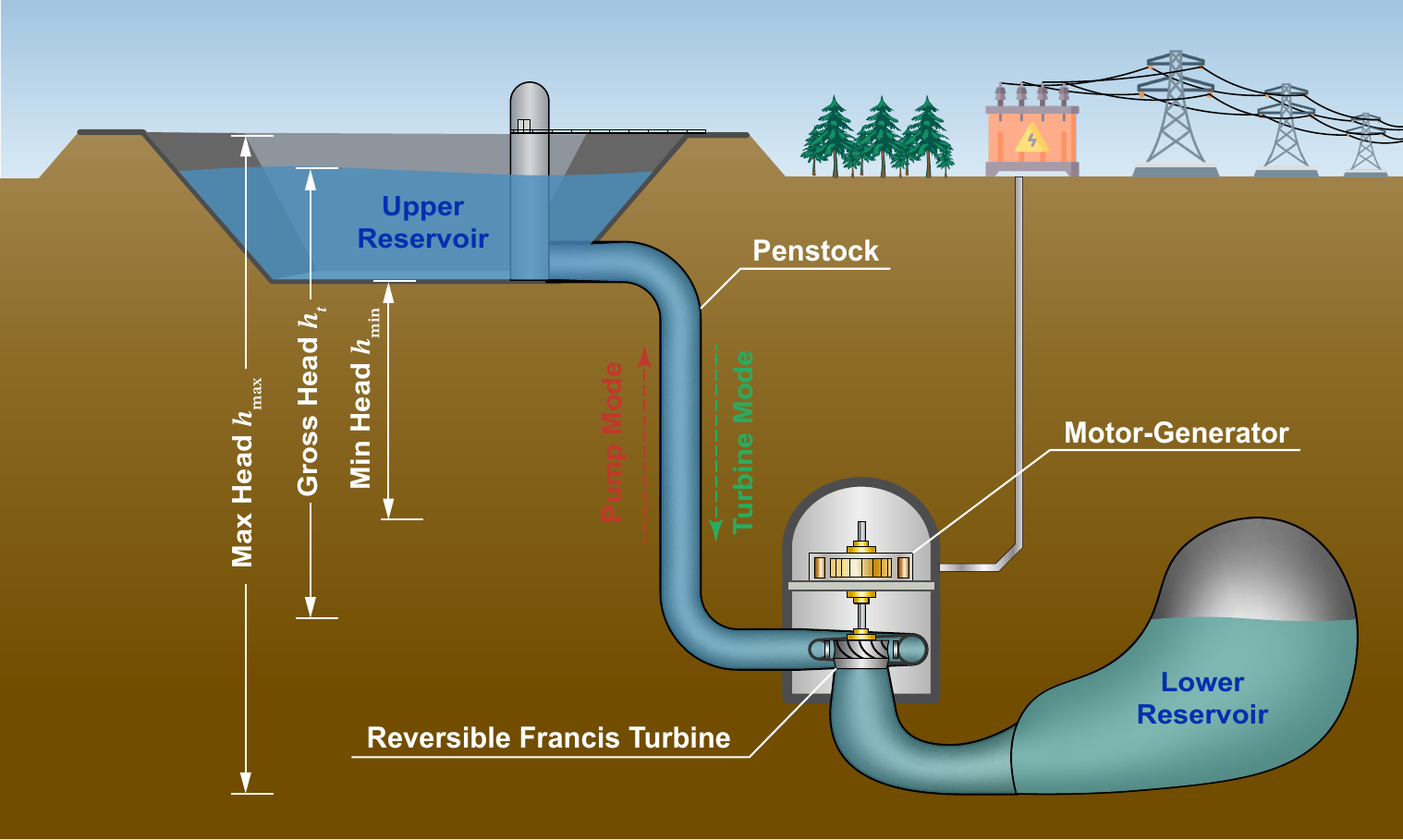}
    \caption{Schematic of the UPHES system showing the reservoirs, reversible Francis turbine, and gross head $h_t$ bounded by $h_{\min}$ and $h_{\max}$.}
    \label{fig:uphes_system}
\end{figure}

Fig.~\ref{fig:uphes_system} illustrates the system configuration. In turbine mode, water descends to generate electricity; in pump mode, electrical power lifts water to the upper reservoir. 

\begin{subequations}
\label{eq:uphes_scheduling_problem}
\begin{gather}
\underset{p_t, z_t^I, z_t^T, z_t^P}{\text{maximize}} \quad \sum_{t=1}^{24} \Delta t \left( p_t \cdot \lambda^{\text{DA}}_t - C_{\text{op}} \cdot p_t^2 \right) \label{eq:objective_complete} \\[2pt]
\text{s.t.} \qquad \sum_{m \in \mathcal{M}} z_t^m = 1, \quad z_t^m \in \{0,1\}, \quad \forall t \in \mathcal{T} \label{eq:mode_sum_binary}\\
p_t = p_t^T + p_t^P,\quad q_t = q_t^T + q_t^P,\quad \forall t \in \mathcal{T} \label{eq:agg_power_flow}\\
 m = I, \ z_t^m = 1 \ \Rightarrow\  p_t = 0,\ q_t = 0, \quad \forall t \in \mathcal{T} \label{eq:idle_indicator}\\
(p_t^m, q_t^m) = 
\begin{cases}
\big(p \in [p_{\min}^m(h_t), p_{\max}^m(h_t)],\ f_m^{\text{UPC}}(p_t^m, h_t)\big), \\
\phantom{(0, 0),} \hspace{0.7em} \text{if } z_t^m = 1, \forall t \in \mathcal{T},\ \forall m \in \{T, P\} \\
(0, 0), \hspace{0.5em} \text{if } z_t^m = 0, \forall t \in \mathcal{T},\ \forall m \in \{T, P\}
\end{cases} 
\label{eq:upc_constraints}\\
v_{\text{low},1} = v_{\text{low}}^{\text{init}} + \Delta t \cdot q_1 \label{eq:volume_initial} \\[6pt]
v_{\text{low},t} = v_{\text{low},t-1} + \Delta t \cdot q_t, \quad \forall t \in \{2,3,\ldots,24\} \label{eq:volume_dynamics} \\
v_{\text{low},t} = f^{\text{vol}}(h_t), \quad \forall t \in \mathcal{T} \label{eq:volume_head} \\
h_{\min} \le h_t \le h_{\max}, \quad \forall t \in \mathcal{T} \label{eq:head_bounds} \\
v_{\text{low},24} \leq v_{\text{low}}^{\text{target}} \label{eq:Target}
\end{gather}
\end{subequations}

Here, $t \in \mathcal{T}=\{1,\dots,24\}$ indexes hourly intervals, and $m \in \mathcal{M}=\{I, T, P\}$ denotes the operating mode (Idle, Turbine, Pump), with binary indicators $z_t^m$. Continuous variables $p_t$, $q_t$, $h_t$, and $v_{\text{low},t}$ represent net power [MW], flow [m$^3$/s], hydraulic head [m], and lower reservoir volume [m$^3$], while $\lambda_t^{\text{DA}}$ and $C_{\text{op}}$ define the market price and cost coefficient.

The objective function \eqref{eq:objective_complete} maximizes the operator's profit over a 24-hour horizon. The operator participates as a price taker in the day-ahead energy market, which clears hourly electricity prices $\lambda_t^{\text{DA}}$ for the following day. The revenue term $p_t \cdot \lambda^{\text{DA}}_t$ captures the electricity production value (positive for turbine operation, negative for pumping). The quadratic operational cost term $C_{\text{op}} \cdot p_t^2$ reflects the nonlinear operational expenses, including mechanical wear, friction losses, and efficiency degradation at higher loading~\cite{yin2023predictive, aburub2019use}. The coefficient $C_{\text{op}}$ effectively aggregates these degradation mechanisms into a scalar operational cost, while $\Delta t$ is the hourly time step.

\subsection{Operational Constraints for Francis Turbine}

\begin{figure}[t!]
    \centering
    \includegraphics[width=\linewidth]{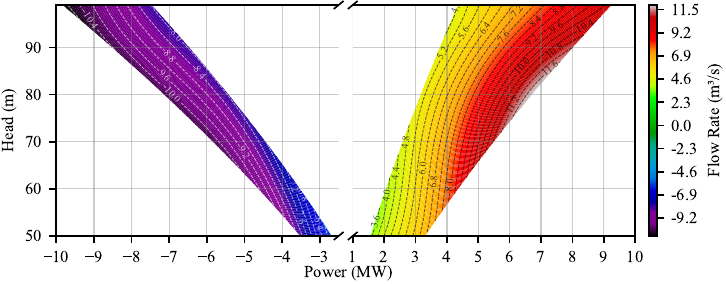}
    \caption{Polynomial fit of the unit performance curves: (left) Pump mode, (right) Turbine mode.}
    \label{fig:UPC_poly_fit_2D}
\end{figure}

Fig.~\ref{fig:UPC_poly_fit_2D} illustrates the unit performance curves (UPCs) obtained from laboratory measurements on a reduced-scale reversible Francis pump–turbine reported by Mercier et~al.~\cite{mercier2017provision} within the SmartWater project~\cite{smartwater_multitel_2022}. The operating space consists of two disjoint manifolds (turbine and pump) separated by an idle transition at zero flow. This discontinuous, non-convex geometry reflects distinct hydraulic regimes: turbine mode extracts energy from descending flow, pump mode lifts water against gravity, and idle mode maintains zero flow during transitions. The mapping between head, power, and flow is consequently non-linear, non-convex, and discontinuous across these operational domains. 
Making an accurate representation of these performance characteristics is essential for optimal scheduling. Francis pump–turbines are employed for their rapid load regulation~\cite{iliev2019variable}, suitability for high-head applications~\cite{wang2020transient}, and established use in utility-scale energy storage~\cite{koritarov2022review}.

To embed the UPCs into an optimization framework, we approximate the experimental data with a bivariate polynomial that maps head and power to discharge. The polynomial degree $d=5$ was selected through evaluation of approximation accuracy versus model complexity. We obtain the mode-dependent surrogate
\begin{equation}
\label{eq:upc-poly}
q_t^{m} = f_m^{\mathrm{UPC}}(p_t,h_t)
=
\sum_{a=0}^{d}\sum_{b=0}^{d-a} c^{m}_{ab}\,p_t^{\,a}\,h_t^{\,b},
\quad m\in\{T,P\},
\end{equation}
where the coefficients $c^{m}_{ab}$ are estimated by least squares. Because turbine and pump samples occupy distinct regions of the $(p_t,h_t)$ plane, \eqref{eq:upc-poly} is fitted separately for each mode. The polynomial approximation achieves high fidelity with $R^2 \geq 0.99988$ for both turbine and pump modes. The projected envelopes are shown in Fig.~\ref{fig:UPC_poly_fit_2D}.

\subsubsection{Mode Selection Constraints}
Constraint~\eqref{eq:mode_sum_binary} ensures mutual exclusivity among operational modes through binary indicators $z_t^m$, at each time $t$. These indicators govern the aggregated power $p_t$ and flow $q_t$ variables defined in~\eqref{eq:agg_power_flow}.

\subsubsection{Idle Mode Constraints}
When idle mode is engaged ($z_t^I=1$), constraint~\eqref{eq:idle_indicator} enforces $p_t=0$ and $q_t=0$, eliminating power exchange and hydraulic flow. Through~\eqref{eq:agg_power_flow}, this condition propagates to both mode-specific variables, while~\eqref{eq:upc_constraints} ensures $(p_t^T, q_t^T) = (0,0)$ and $(p_t^P, q_t^P) = (0,0)$ when the respective mode indicators are inactive.

\subsubsection{Active Mode Constraints}
For active modes $m \in \{T, P\}$ with $z_t^m=1$, constraint~\eqref{eq:upc_constraints} bounds power within $[p_{\min}^m(h_t), p_{\max}^m(h_t)]$, and enforces flow rate $q_t^m = f_m^{\text{UPC}}(p_t^m, h_t)$ via ~\eqref{eq:upc-poly}. When $z_t^m=0$, both $p_t^m$ and $q_t^m$ are set to zero, decoupling the inactive mode from system dynamics. The head-dependent power boundaries ensure operation remains within the hydraulically admissible envelope, as operating outside this range is hazardous: excessive flow can trigger damaging cavitation (rapid implosion of gaseous cavities) in the Francis turbine, whereas insufficient flow can lead to self-excited vibrations and severe erosion~\cite{pannatier2010optimisation}.

\subsection{Volume Dynamics Constraints}
\label{sec:volume_constraints}

The lower reservoir volume evolves according to volume conservation, where the rate of volume change equals the net water flow: $\frac{dv_{\text{low}}}{dt}=q(t)$. Constraints~\eqref{eq:volume_dynamics} and~\eqref{eq:volume_initial} discretize this continuous balance using forward Euler integration with time step $\Delta t = 3600$~s, linking hourly flow decisions to cumulative storage levels throughout the scheduling horizon.

The nonlinear volume–head coupling arises from the modeling assumptions on reservoir geometries that deliberately introduce nonlinearity and nonconvexity into the optimization problem. These geometric configurations capture the computational challenges inherent in real UPHES systems while maintaining analytical tractability. Following the frustum-based storage-elevation modeling approach~\cite{liu2022low}, the upper reservoir is represented as a frustum with average slope $s$, yielding the cubic volume-head relationship
\begin{align}
v_{\text{up}}(h_{\text{up}}) = \pi r_{\text{base}}^2 h_{\text{up}} + \pi s\, r_{\text{base}} h_{\text{up}}^2 + \frac{\pi s^2}{3} h_{\text{up}}^3, \label{eq:v_up_cubic}
\end{align}
where $r_{\text{base}}$
 is the base radius of the frustum. For the lower reservoir, various underground configurations have been proposed for UPHES systems, including room-and-pillar layouts and interconnected cavern networks~\cite{allen1984underground}. Following established modeling approaches, we represent the lower reservoir as $n$ spherical mine pits, yielding
\begin{align}
v_{\text{low}}(h_{\text{low}}) = n\pi R h_{\text{low}}^2 - \frac{n\pi}{3} h_{\text{low}}^3,\label{eq:v_low_cubic}
\end{align}
where $R$
 is the sphere radius. The gross head is given as
\begin{align}
h = f_{\text{up}}^{\text{vol}\,-1}(v_{\text{up}}) - f_{\text{low}}^{\text{vol}\,-1}(v_{\text{low}}), \label{eq:head_vol_relation}
\end{align}
subject to $v_{\text{up}} + v_{\text{low}} = v_{\text{total}}$, where $f_{\text{up}}^{\text{vol}\,-1}$ and $f_{\text{low}}^{\text{vol}\,-1}$ denote the inverse mappings obtained by solving the cubic equations \eqref{eq:v_up_cubic} and (\ref{eq:v_low_cubic}), respectively.

The hydraulic head is constrained within its feasible operating range as enforced by~\eqref{eq:head_bounds}. This condition ensures that the computed head levels remain physically realistic and consistent with the reservoir geometry throughout the scheduling horizon.

Target constraint \eqref{eq:Target} ensures that the final lower reservoir volume $v_{\text{low},t=24}$ does not exceed the target $v_{\text{low}}^{\text{target}}$, preserving long-term water balance and operational consistency.

\section{Preliminaries}
\label{sec:preliminaries}

The MINLP described in Section~II poses computational challenges due to polynomial UPC mappings~\eqref{eq:upc-poly}, cubic volume-head relations~\eqref{eq:v_up_cubic} and~\eqref{eq:v_low_cubic}, and head-dependent power boundaries. This section presents two tractable piecewise reformulations of these nonlinear relationships that retain the binary mode variables from~\eqref{eq:mode_sum_binary}. These formulations serve as baselines for evaluating the DFL methodology in Section~\ref{sec:methodology}.

\subsection{Global Linearization Approach}

Global linearization approximates each nonlinear relationship with a  affine function fitted over the entire operational domain, trading accuracy for computational efficiency. For the UPC mappings in~\eqref{eq:upc-poly}, we replace the polynomial with
\begin{align}
q_t^m = 
\begin{bmatrix} 
p_t^m & h_t & 1 
\end{bmatrix}
\boldsymbol{\alpha}_m,
\quad \forall t \in \mathcal{T},\ m \in \{T,P\}
\label{eq:upc_linear}
\end{align}
where $\boldsymbol{\alpha}_m \in \mathbb{R}^3$ is obtained by least-squares fitting on experimental data. The power boundaries are linearized as
\begin{align}
p_{\min}^m(h_t) &= 
\begin{bmatrix} 
h_t & 1 
\end{bmatrix}
\boldsymbol{\beta}_m^{\min}, 
\quad
p_{\max}^m(h_t) = 
\begin{bmatrix} 
h_t & 1 
\end{bmatrix}
\boldsymbol{\beta}_m^{\max}, \label{eq:pminmax_linear}
\end{align}
for $m \in \{T,P\}$. The volume-head relationship~\eqref{eq:volume_head} becomes
\begin{align}
h_t = 
\begin{bmatrix} 
v_{\text{low},t} & 1 
\end{bmatrix}
\boldsymbol{\delta}, 
\quad \forall t \in \mathcal{T}
\label{eq:vh_linear}
\end{align}
where $\boldsymbol{\delta} \in \mathbb{R}^2$. Idle-mode constraints~\eqref{eq:idle_indicator} are enforced via big-$M$ formulations. While computationally efficient, the single-plane approximation introduces systematic errors in high-curvature regions, motivating the adaptive local linearization strategy developed in Section~\ref{sec:methodology}.

\subsection{Piecewise Bilinear Approximation with SOS2 Constraints}

This second method enhances the fidelity of the UPC approximation while retaining a tractable formulation. We discretize the function space on the variable grid and enforce interpolation only between adjacent grid vertices. The resulting interpolant is a bilinear hyperbolic paraboloid. A Special Ordered Set of type 2 (SOS2) constraint~\cite{beale1970special} is employed to ensure that the convex‑combination weights are non‑zero only for two neighboring vertices in each cell~\cite{beale1976global}. This approach is grounded in modeling of non‑separable multivariate functions in a mixed-integer manner~\cite{huchette2019combinatorial,barmann2023piecewise}.

For the volume-head, SOS2 variables $\Omega_{t,i}^{\text{vh}} \in [0,1]$ satisfy
\begin{equation}
\sum_{i \in \mathcal{I}_h} \Omega_{t,i}^{\text{vh}} = 1, \quad
(h_t, v_{\text{low},t}) = \sum_{i \in \mathcal{I}_h} \Omega_{t,i}^{\text{vh}} (h_i^{\text{sample}}, v_{\text{low},i}^{\text{sample}}),
\label{eq:sos2_vh}
\end{equation}
where at most two adjacent $\Omega_{t,i}^{\text{vh}}$ are nonzero. For UPC mappings, auxiliary variables $\Omega_{t,i,j}^m \in [0,1]$ satisfy
\begin{equation}
\sum_{j \in \mathcal{I}_p^m} \Omega_{t,i,j}^m = z_t^m \cdot \Omega_{t,i}^{\text{vh}}, \quad
\sum_{i \in \mathcal{I}_h}\sum_{j \in \mathcal{I}_p^m} \Omega_{t,i,j}^m = z_t^m,
\label{eq:sos2_mode}
\end{equation}
with power and flow reconstructed as
\begin{equation}
(p_t^m, q_t^m) = \sum_{i \in \mathcal{I}_h}\sum_{j \in \mathcal{I}_p^m} \Omega_{t,i,j}^m (p_{i,j}^{m,{\text{sample}}}, q_{i,j}^{m,{\text{sample}}}).
\label{eq:sos2_power_flow}
\end{equation}
This formulation improves accuracy over global linearization at the cost of $O(N_h \cdot N_p^T + N_h \cdot N_p^P)$ auxiliary variables, where $N_h$ denotes the number of head discretization points and $N_p^m$ represents the number of power discretization points for mode $m \in \{T, P\}$. The computational complexity scales linearly with grid resolution, requiring $24(N_h \cdot N_p^T + N_h \cdot N_p^P)$ SOS2 interpolation weights $\Omega_{t,i,j}^m$ over the planning horizon.

\section{Methodology}
\label{sec:methodology}

\begin{figure}[t!]
\centering
\includegraphics[width=\columnwidth]{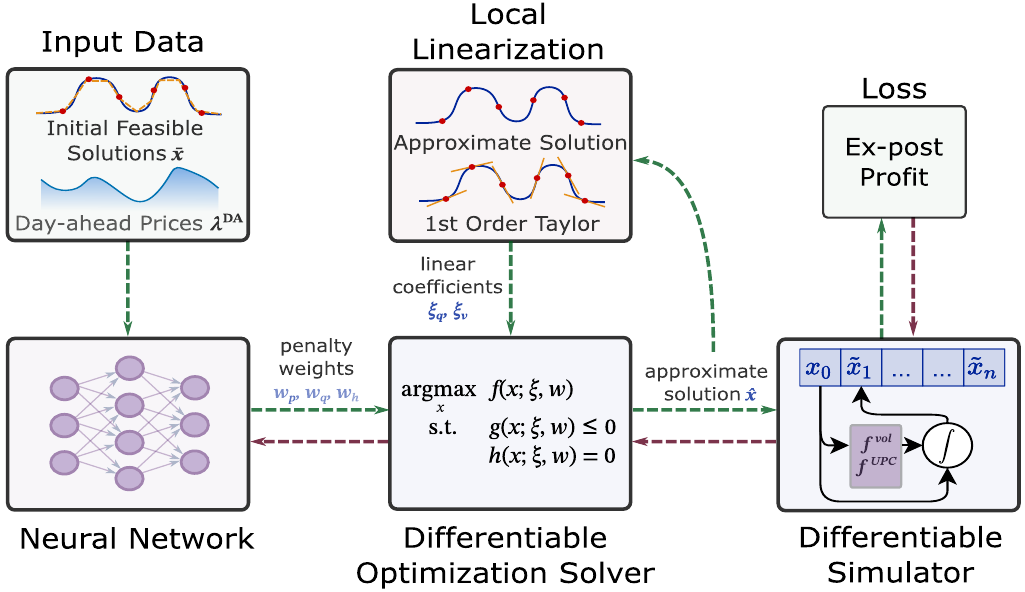}
\caption{Decision-focused learning for UPHES day-ahead scheduling. Green arrows represent forward pass, while red arrows represent backward pass.}
\label{fig:methodology_framework}
\end{figure}

This section describes a DFL framework for computationally efficient and accurate UPHES day-ahead scheduling. As illustrated in Fig.~\ref{fig:methodology_framework} and detailed in Algorithm~\ref{alg:dfl_scheduling}, the methodology integrates four components: (\textit{i}) a neural penalty predictor (Section~\ref{subsec:penalty_prediction}) that generates adaptive regularization weights to establish trust regions ensuring linearization validity, (\textit{ii}) a local linearization layer (Section~\ref{subsec:local_linearization}) that constructs first-order Taylor approximations converting nonlinear constraints into convex form, (\textit{iii}) a differentiable convex optimizer (Section~\ref{subsec:diff_optimizer}) that iteratively refines schedules through penalty-guided quadratic programming while preserving mode consistency, and (\textit{iv}) a physical simulator (Section~\ref{subsec:simulation}) that enforces true nonlinear dynamics to validate feasibility and quantify the economic performance used as training signal. 
This end-to-end differentiable DFL architecture enables optimization of economic performance across a range of price scenarios and can be used as a computationally tractable refiner of initial solutions obtained either from a baseline MIQP with global linearization (Section~\ref{sec:preliminaries}) or historical schedules.

\begin{algorithm}[t!]
\caption{End-to-End Decision-Focused Learning}
\label{alg:dfl_scheduling}
\begin{algorithmic}[1]
\Statex \textbf{Input:} Training set $\mathcal{D} = \{(\boldsymbol{\lambda}_i^{\text{DA}}, \bar{\mathbf{x}}_i)\}_{i=1}^N$
\Statex \textbf{Output:} Trained parameters $\theta^*$
\vspace{1pt}
\Statex \hrule
\State Initialize $\theta^{(0)}$ 
\For{epoch $e = 1$ to $E$}
    \For{each sample $(\boldsymbol{\lambda}_i^{\text{DA}}, \bar{\mathbf{x}}_i)$ in $\mathcal{D}$}
        \State $\mathbf{w}_i^{(0)} \gets \exp(\mathcal{N}_{\theta^{(e)}}([\boldsymbol{\lambda}_i^{\text{DA}}, \bar{\mathbf{x}}_i]))$, $\hat{\mathbf{x}}_i^{(0)} \gets \bar{\mathbf{x}}_i$ 
        \For{$k = 0$ to $K-1$}
            \State $\mathbf{w}_i^{(k)} \gets \gamma^k \mathbf{w}_i^{(0)}$
            \State Linearize $f^{\text{UPC}}$ and $f^{\text{vol}}$ at $\hat{\mathbf{x}}_i^{(k)}$ to obtain $\boldsymbol{\xi}_i^{(k)}$
            \State Solve QP~\eqref{eq:opt_objective_penalty} with $\boldsymbol{\xi}_i^{(k)}$, $\mathbf{w}_i^{(k)}$ to get $\hat{\mathbf{x}}_i^{(k+1)}$
        \EndFor
        \State $\tilde{\mathbf{x}}_i \gets \textsc{Simulate}(\hat{\mathbf{x}}_i^{(K)})$ using Algorithm~\ref{alg:simulation}
        \State $\mathcal{L}_i \gets -\Pi_i(\tilde{\mathbf{x}}_i)$ via~\eqref{eq:expost_profit}
    \EndFor
    \State $\mathcal{L}(\theta^{(e)}) \gets \frac{1}{N}\sum_{i=1}^N \mathcal{L}_i$
    \State Update parameters: $\theta^{(e+1)} \gets \theta^{(e)} - \eta \nabla_\theta \mathcal{L}(\theta^{(e)})$
\EndFor
\State \Return $\theta^*$
\end{algorithmic}
\end{algorithm}

\subsection{Neural Penalty Prediction}
\label{subsec:penalty_prediction}

The neural penalty predictor estimates time-varying regularization weights that guide the linearization of noisy initial solutions. Given the day-ahead price vector $\boldsymbol{\lambda}^{\text{DA}} = [\lambda_1^{\text{DA}}, \ldots, \lambda_{24}^{\text{DA}}]^\top$ and initial noisy trajectories $\bar{\mathbf{x}} = \{\bar{p}_t, \bar{q}_t, \bar{h}_t\}_{t \in \mathcal{T}}$ from, e.g., a baseline MIQP solver, the network constructs a feature matrix $\mathbf{X} \in \mathbb{R}^{24 \times 4}$ where each row contains $[\lambda_t^{\text{DA}}, \bar{p}_t, \bar{q}_t, \bar{h}_t]$. This representation captures both market signals and the physical states of the reservoir. 

An LSTM network processes the sequential input to capture temporal dependencies in pricing patterns and operating schedules. The recurrent architecture recognizes multi-hour pumping and generation cycles typical of energy storage dispatch. The final hidden state at each time step passes through a linear projection to produce log-domain penalty weights

\begin{equation}
[\log \mathbf{w}_p, \log \mathbf{w}_q, \log \mathbf{w}_h] = \text{Linear}(\text{LSTM}(\mathbf{X})),
\label{eq:network_output_vector_explicit}
\end{equation}

which are exponentiated and bounded to prescribed intervals to ensure numerical stability and prevent penalty weights from dominating the objective. Log-domain parameterization guarantees strictly positive penalty weights while accommodating the wide dynamic range across different operating conditions. \
The optimal trust region size at each time step depends on the local curvature of the nonlinear UPC and volume-head functions at the current operating point, which varies across the horizon and across scenarios. A  static weight triple calibrated offline cannot encode this variation, motivating a learned, scenario-adaptive predictor (validated empirically in Section~\ref{sec:results}).

\subsection{Local Linearization}
\label{subsec:local_linearization}

The recursive linearization process constructs first-order Taylor approximations of the nonlinear UPC and volume-head relationships around the current operating point. At iteration $k$ within the inner loop of Algorithm~\ref{alg:dfl_scheduling}, the operating point $\hat{\mathbf{x}}^{(k)} = \{\hat{p}_t^{(k)}, \hat{q}_t^{(k)}, \hat{h}_t^{(k)}\}_{t \in \mathcal{T}}$ defines the expansion point for local linear approximations. 

Following this approximation, the UPC mapping~\eqref{eq:upc-poly} becomes
\begin{equation}
q_t \approx \begin{bmatrix} p_t & h_t & 1 \end{bmatrix} \boldsymbol{\xi}_{q,t}^{(k)},
\label{eq:taylor_flow}
\end{equation}
for each time step $t$ and active mode $m$ where
\begin{equation}
\boldsymbol{\xi}_{q,t}^{(k)} = \begin{bmatrix}
\left. \frac{\partial f_m^{\text{UPC}}}{\partial p} \right|_{(\hat{p}_t^{(k)}, \hat{h}_t^{(k)})} \\[4pt]
\left. \frac{\partial f_m^{\text{UPC}}}{\partial h} \right|_{(\hat{p}_t^{(k)}, \hat{h}_t^{(k)})} \\[4pt]
f_m^{\text{UPC}}(\hat{p}_t^{(k)}, \hat{h}_t^{(k)}) - \nabla f_m^{\text{UPC}}(\hat{p}_t^{(k)}, \hat{h}_t^{(k)}) \begin{bmatrix} \hat{p}_t^{(k)} \\ \hat{h}_t^{(k)} \end{bmatrix}
\end{bmatrix}.
\label{eq:xi_flow}
\end{equation}
Similarly, the volume-head coupling~\eqref{eq:volume_head} is linearized as
\begin{equation}
v_{\text{low},t} \approx \begin{bmatrix} h_t & 1 \end{bmatrix} \boldsymbol{\xi}_{v,t}^{(k)},
\label{eq:taylor_volume}
\end{equation}
with coefficient vector $\boldsymbol{\xi}_{v,t}^{(k)} \in \mathbb{R}^2$ defined by the first-order Taylor expansion of $f_{\text{low}}^{\text{vol}}$ around $\hat{h}_t^{(k)}$:
\begin{equation}
\boldsymbol{\xi}_{v,t}^{(k)} = \begin{bmatrix}
\left. \frac{d f_{\text{low}}^{\text{vol}}}{d h} \right|_{\hat{h}_t^{(k)}} \\[4pt]
f_{\text{low}}^{\text{vol}}(\hat{h}_t^{(k)}) - \left. \frac{d f_{\text{low}}^{\text{vol}}}{d h} \right|_{\hat{h}_t^{(k)}} \hat{h}_t^{(k)}
\end{bmatrix}.
\label{eq:xi_volume}
\end{equation}

All partial derivatives are computed via automatic differentiation, ensuring accurate gradient information of the polynomial UPC~\eqref{eq:upc-poly} and cubic volume-head functions. When $|\hat{p}_t^{(k)}| < \epsilon$, indicating idle mode, $\boldsymbol{\xi}_{q,t}^{(k)}=0$ enforces $q_t = 0$. 

\subsection{Differentiable Optimization}
\label{subsec:diff_optimizer}

Using the linearization coefficients computed in the previous subsection, the optimization layer solves a convex quadratic program at each iteration of the inner loop. The QP reformulates the original UPHES scheduling problem~\eqref{eq:uphes_scheduling_problem} by incorporating adaptive regularization penalties while replacing nonlinear constraints with their local linearizations.
The modified objective function becomes
\begin{align}
\max_{p,q,h} \quad 
& \Delta t \sum_{t \in \mathcal{T}} \left( \lambda_t^{\text{DA}} p_t - C_{\text{op}} p_t^2 \right) - \sum_{t \in \mathcal{T}} \big[ w_{p,t} (p_t - \hat{p}_t^{(k)})^2 \notag \\
& \quad + w_{h,t} (h_t - \hat{h}_t^{(k)})^2 + w_{q,t} (q_t - \hat{q}_t^{(k)})^2 \big], \label{eq:opt_objective_penalty} 
\end{align}
where the quadratic penalty terms implement a soft trust region around the linearization point $\hat{\mathbf{x}}^{(k)}$. Because the Taylor expansions~\eqref{eq:taylor_flow}--\eqref{eq:taylor_volume} are accurate only near the expansion point, the weights control how far the QP solution may deviate to preserve  linearization accuracy, balancing it against profit maximization within a single concave objective.

The nonlinear UPC~\eqref{eq:upc_constraints} and volume-head~\eqref{eq:volume_dynamics} relationships are replaced by their local linearizations~\eqref{eq:taylor_flow} and~\eqref{eq:taylor_volume}. Mode constraints prevent transitions during linearization
\begin{align}
p_t \in \begin{cases}
[p_{\min}^T(h_t), p_{\max}^T(h_t)] & \text{if } \hat{p}_t^{(k)} > 0,\\
[p_{\min}^P(h_t), p_{\max}^P(h_t)] & \text{if } \hat{p}_t^{(k)} < 0,\\
\{0\} & \text{if } \hat{p}_t^{(k)} = 0,
\end{cases} \\
    q_t = 0 \quad \text{ if } \hat{p}_t^{(k)} = 0. \label{eq:opt_mode}
\end{align}
All remaining constraints~\eqref{eq:head_bounds}--\eqref{eq:Target} remain unchanged. This mode-locking strategy eliminates integer variables, transforming the MINLP into a tractable convex QP while maintaining the discrete structure inherited from the initial solution.

The convex QP is implemented using CVXPY with CVXPYLayers~\cite{cvxpylayers2019}, which applies the implicit function theorem to compute gradients of optimal primal variables with respect to penalty weights without differentiating through iterative solver steps.

After solving, the updated trajectories $\hat{\mathbf{x}}^{(k+1)} = \{\hat{p}_t^{(k+1)}, \hat{q}_t^{(k+1)}, \hat{h}_t^{(k+1)}\}_{t \in \mathcal{T}}$ replace the previous iterate for the next inner loop iteration. Penalty weights are scaled by growth factor $\gamma^k$ at each iteration, progressively tightening constraints to encourage convergence through an annealing effect where early iterations explore broader regions while later iterations refine solutions locally.

\subsection{Simulation Layer and Ex-Post Profit Evaluation}
\label{subsec:simulation}

The simulation layer validates physical feasibility by enforcing the original nonlinear UPC model~\eqref{eq:upc-poly} and volume-head relationship~\eqref{eq:head_vol_relation}. Given the optimized power trajectory $\hat{\mathbf{x}}^{(K)} = \{\hat{p}_t^{(K)}\}_{t \in \mathcal{T}}$ from the final iteration, the simulator computes actual flow rates $\tilde{\mathbf{x}} = \{\tilde{p}_t, \tilde{q}_t, \tilde{h}_t, \tilde{v}_{\text{low},t}\}_{t \in \mathcal{T}}$ by evaluating the polynomial UPC at realized head values, which are updated sequentially according to reservoir mass balance
\begin{align}
\tilde{v}_{\text{low},0} &= v_{\text{low}}^{\text{init}} \label{eq:sim_v0}\\
\tilde{v}_{\text{low},t} &= \tilde{v}_{\text{low},t-1} + \tilde{q}_t \cdot \Delta t, \quad \forall t \in \mathcal{T} \label{eq:sim_vt}\\
\tilde{h}_t &= f_{\text{low}}^{\text{vol},-1}(\tilde{v}_{\text{low},t}), \quad \forall t \in \mathcal{T}. \label{eq:sim_ht}
\end{align}
Power outputs are clamped to head-dependent feasibility bounds to prevent UPC envelope violations. If predicted volume exceeds upper reservoir capacity or falls below the minimum admissible level, the unit is forced into idle mode. This clamping and mode switching logic ensures physical realizability even when the optimization layer produces infeasible suggestions due to approximation errors. \
For backpropagation, these non-smooth operations use a straight-through estimator~\cite{bengio2013estimating}: the forward pass applies the hard physical correction while the backward pass treats it as the identity, preserving gradient flow through the simulation layer. Algorithm~\ref{alg:simulation} details this procedure.

\begin{algorithm}[t!]
\caption{Physical Simulation Layer}
\label{alg:simulation}
\begin{algorithmic}[1]
\Statex \textbf{Input:} Optimized power schedule $\hat{\mathbf{p}} = \{\hat{p}_t\}_{t \in \mathcal{T}}$
\Statex \textbf{Output:} Feasible simulated trajectories $\{\tilde{\mathbf{p}}, \tilde{\mathbf{q}}, \tilde{\mathbf{h}}, \tilde{\mathbf{v}}\}$
\vspace{1pt}
\Statex \hrule
\State Initialize: $\tilde{v}_{\text{low}} \gets v_{\text{low}}^{\text{init}}$, $\tilde{h} \gets h^{\text{init}}$
\For{$t = 1$ to $24$}
    \State Determine mode from $\hat{p}_t$ (idle, turbine, or pump)
    \State Clamp $\tilde{p}_t$ to $[p_{\min}^m(\tilde{h}_t), p_{\max}^m(\tilde{h}_t)]$
    \State Evaluate flow: $\tilde{q}_t \gets f_m^{\text{UPC}}(\tilde{p}_t, \tilde{h}_t)$
    \State Update volume: $\tilde{v}_{\text{low},t} \gets \tilde{v}_{\text{low},t-1} + \tilde{q}_t \cdot \Delta t$
    \If{$\tilde{v}_{\text{low},t} > v_{\max}$ or $\tilde{v}_{\text{low},t} < v_{\min}$}
        \State Force idle: $\tilde{p}_t \gets 0$, $\tilde{q}_t \gets 0$, $\tilde{v}_{\text{low},t} \gets \tilde{v}_{\text{low},t-1}$
    \EndIf
    \State Update head: $\tilde{h}_{t+1} \gets f_{\text{low}}^{\text{vol},-1}(\tilde{v}_{\text{low},t})$
\EndFor
\State \Return $\{\tilde{\mathbf{p}}, \tilde{\mathbf{q}}, \tilde{\mathbf{h}}, \tilde{\mathbf{v}}\}$
\end{algorithmic}
\end{algorithm}

The ex-post profit metric quantifies economic performance by accounting for revenue, operational costs, and penalties for constraint violations
\begin{align}
\Pi= \Delta t \sum_{t \in \mathcal{T}} \left( \lambda_t^{\text{DA}} \tilde{p}_t - C_{\text{op}} \tilde{p}_t^2 \right) - \text{SI} - \text{Vol},
\label{eq:expost_profit}
\end{align}
where the system imbalance penalty $\text{SI}$ accounts for deviations between optimized and simulated power outputs through asymmetric pricing: energy production shortages are penalized at twice the day-ahead price while surpluses are compensated only at half the day-ahead price. \
This asymmetry reflects Belgian single-imbalance pricing, where the marginal imbalance price for shortfalls is  above the day-ahead price and the downward counterpart is  below it~\cite{toubeau2019chance, favaro_multi-fidelity_2024}. The volume penalty $\text{Vol}$ converts any excess water remaining beyond the target reservoir \eqref{eq:Target} level into equivalent energy value priced at the median day-ahead rate,\
 following the established water-value concept in hydropower economics~\cite{pereira1991multi}. No penalty is assessed when the terminal volume constraint is satisfied.

\subsection{Training Procedure}
\label{subsec:train_procedure}

Our goal is to train the neural penalty predictor offline using noisy MIQP solutions generated by perturbing baseline piecewise SOS2 optimization results. Perturbations are applied to power schedule variables $\{p_t\}_{t \in \mathcal{T}}$ while preserving operational modes $\{m_t \in \{I, T, P\}\}_{t \in \mathcal{T}}$ from the baseline solution. Perturbation magnitudes are uniformly sampled from intervals proportional to head-dependent capacity ranges $[p_{\min}^m(h_t), p_{\max}^m(h_t)]$, spanning noise levels from 10\% to 80\% as well as random sampling. Physically consistent flow rates $\{q_t\}_{t \in \mathcal{T}}$ are recovered by evaluating the polynomial UPC model~\eqref{eq:upc-poly} at the perturbed power values, and head trajectories $\{h_t\}_{t \in \mathcal{T}}$ are obtained by integrating volume dynamics~\eqref{eq:volume_dynamics} with the volume-head relationship~\eqref{eq:volume_head}. This generates training pairs $\{(\boldsymbol{\lambda}_i^{\text{DA}}, \bar{\mathbf{x}}_i) \mapsto \Pi_i\}_{i=1}^N$ where $\bar{\mathbf{x}}_i = \{\bar{p}_{i,t}, \bar{q}_{i,t}, \bar{h}_{i,t}\}_{t \in \mathcal{T}}$.

The framework is trained per representative day. Each of the 19 representative price profiles (Section~\ref{sec:data}) yields a separate penalty predictor, trained only on perturbations of that day's own MIQP-PW reference schedule. Per day we generate 9 perturbed variants, namely 8 relative noise levels (10\% to 80\%) plus one random-sampling variant, giving $19 \times 9 = 171$ training instances in total. The headline DFL results use the random-sampling variant, which is the most corrupted initialization. Model selection is performed within each day's own perturbed data. Leakage cannot arise: no data from any other day enters a given day's model, and the clean reference schedule on which a day is evaluated is never one of its perturbed training instances.

The loss function maximizes ex-post profit~\eqref{eq:expost_profit} obtained from the simulation layer after iterative linearization
\begin{equation}
\mathcal{L}(\theta) = -\frac{1}{N} \sum_{i=1}^N \Pi_i(\theta),
\label{eq:loss}
\end{equation}
where $\theta$ denotes network parameters. Training uses the Adam optimizer with plateau-based learning rate scheduling, early stopping, and gradient clipping. Table~\ref{tab:repro} summarizes the architecture, penalty bounds, iteration count, solver tolerances, and training settings used for all reported DFL results.

\begin{table}[t!]
\centering
\caption{Reproducibility Settings for the DFL Framework}
\label{tab:repro}
\footnotesize
\setlength{\tabcolsep}{4pt}
\renewcommand{\arraystretch}{0.95}
\begin{tabular}{@{}ll@{}}
\toprule
Component & Setting \\
\midrule
Architecture & LSTM, 3 layers, hidden 128, dropout 0.2 \\
Input / output & 4 features ($\lambda^{\text{DA}}, p, q, h$) / 3 log-weights \\
Bounds $w_p,w_q,w_h$ & $[0.1,3]$, $[0.001,0.2]$, $[0.01,5]$ \\
Init.\ $w_p,w_q,w_h$ & 0.6, 0.02, 0.1; growth 1.5/iter. \\
Iterations $K$ & 7 \\
QP solver & ECOS (CVXPYlayers), $\le 2\times10^{5}$ it., tol.\ $10^{-5}$ \\
Optimizer & Adam, lr $10^{-3}$, plateau LR (0.5, pat.\ 5) \\
Epochs / patience / clip & 500 / 20 / max-norm 1.0 \\
\bottomrule
\end{tabular}
\end{table}

\section{Results and Discussion}
\label{sec:results}

This section validates the proposed DFL framework through experiments on a realistic UPHES site in Belgium using actual electricity price scenarios presented in Section~\ref{sec:data}. Performance evaluation compares DFL methods against MIQP benchmarks from Section~\ref{sec:preliminaries}, and ablation studies are conducted to isolate architectural component contributions. Ex-post profit serves as the primary performance metric, computed by simulating optimized schedules under original system dynamics with realistic market penalties for constraint violations. 

\subsection{Data Description}
\label{sec:data}

The case study considers a 24-hour day-ahead scheduling problem with hourly time steps. As illustrated in Fig.~\ref{fig:uphes_system}, the UPHES plant features hydraulic head varying between $h_{\min} = 50$~m and $h_{\max} = 99$~m. As introduced in Section~\ref{sec:volume_constraints}, the upper reservoir exhibits a frustum geometry with an average slope $s=1.8$, while the lower reservoir consists of $n=100$ interconnected mine pits. Both reservoirs have maximum storage capacities of 588,000~m$^3$ and are initialized at 50\% capacity, with the terminal constraint enforcing $v_{\text{low}}^{\text{target}} = 294,000$~m$^3$ to ensure cyclical operation.
Day-ahead electricity prices $\lambda_t^{\text{DA}}$ are based on representative price scenarios in 2024, provided by Belgian transmission system operator Elia~\cite{EliaDayAheadReferencePrice}. The k-medoids algorithm is preferred over k-means to avoid creating fictitious average profiles. Using k-medoids clustering with three fixed extreme medoids (highest mean price, highest volatility, lowest mean price), we identify 19 representative daily price profiles. These 19 days form the training and validation set: each day's MIQP-PW reference schedule is perturbed to train and select that day's model, as described in Section~\ref{subsec:train_procedure}, and each day is then evaluated on its own clean reference price profile. An additional 19 held-out days, disjoint from the 19 representative days, are drawn from the remaining 346 days of 2024 as a test set for MIP-free deployment: three extremes (highest mean, highest volatility, lowest mean) are fixed, and the other 16 are sampled by nearest-neighbor price-similarity matching to the representative medoids. The operational cost coefficient is set to $C_{\text{op}} = 0.4$ EUR/MW$^2$. The optimization is implemented using CVXPYlayers 0.1.6 with Gurobi 12.0.2 on an Intel Core Ultra 9 275HX Processor with 64 GB RAM running Windows 11. The MIP gap tolerance is set to 1\% with a time limit of 3600~s.

\subsection{Benchmark Approximation Quality}

Before evaluating the DFL framework, we characterize the approximation accuracy of the baseline MIQP methods introduced in Section~\ref{sec:preliminaries}. Table~\ref{tab:linearization_errors} shows approximation fidelity using four metrics evaluated on 480 operating points sampled from preliminary MIQP solutions across the \
19 representative price scenarios. The metrics compare linearization predictions against evaluations using the true polynomial UPC~\eqref{eq:upc-poly} and cubic volume-head relationships~\eqref{eq:v_up_cubic},~\eqref{eq:v_low_cubic}.

\begin{table}[t!]
\centering
\caption{Benchmark Approximation Error Analysis}
\label{tab:linearization_errors}
\small
\begin{tabular}{@{}cccccc@{}}
\toprule
Function & Method & Mean & Max & MAPE & $R^2$ \\
 & & (\%) & (\%) & (\%) & \\
\midrule
\multirow{2}{*}{$f_m^{\text{UPC}}$} 
 & Piecewise Bilinear & 0.16 & 4.09 & 0.20 & 1.0000 \\
 & Global Linear & 4.36 & 18.31 & 5.78 & 0.9913 \\
\midrule
\multirow{2}{*}{$f^{\text{vol}}$} 
 & Piecewise Bilinear & 0.16 & 1.79 & 0.17 & 0.9999 \\
 & Global Linear & 15.98 & 18.08 & 16.08 & 0.4772 \\
\bottomrule
\end{tabular}
\end{table}

The piecewise bilinear approximation achieves MAPE below 0.21\% with near-perfect correlation ($R^2 \geq 0.9999$), while global linearization exhibits MAPE of 5.78\% for UPC and 16.08\% for volume-head mappings. The observed degradation in volume-head approximation accuracy stems from cubic geometry that cannot be captured by single-plane fits.

\subsection{Performance of Proposed Methods and Benchmarks}
\label{subsec:main_results}

\begin{figure*}[t!]
    \centering
    \includegraphics[width=\textwidth]{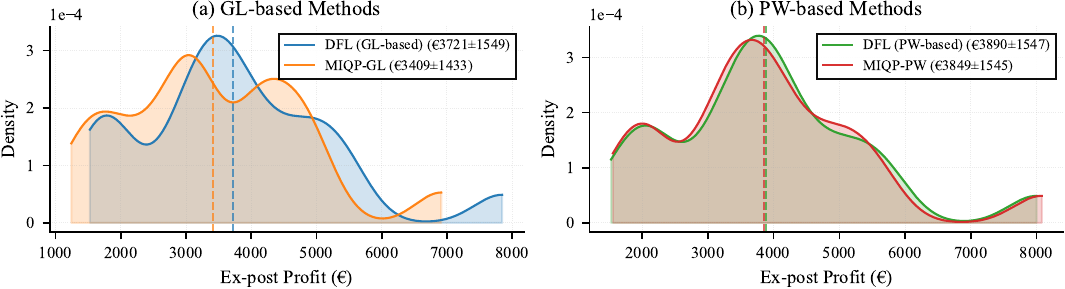} 
    \caption{Density distribution of ex-post profit comparing DFL methods against MIQP benchmarks across 19 representative price scenarios. (a)~GL-based methods, (b)~PW-based methods. Vertical dashed lines indicate mean values. Densities estimated via kernel density estimation with Gaussian kernel (Scott's rule bandwidth, 0.5 scaling factor).}

    \label{fig:profit_density_main}
\end{figure*}

The performance of the proposed DFL framework is evaluated against MIQP benchmarks. Four methodologies are compared: piecewise-based DFL (PW-based DFL) employs piecewise bilinearization within the recursive linearization, global linearization-based DFL (GL-based DFL) utilizes global linear approximations, while benchmark methods MIQP-PW and MIQP-GL represent piecewise SOS2 bilinear and global linear formulations solved through mixed-integer optimization.

Training data for the DFL framework are constructed following the methodology described in Section~\ref{subsec:train_procedure}. Noise-injected MIQP baselines exhibit monotonic performance degradation, with MIQP-PW declining from 3837 EUR at 10\% noise to 3365 EUR under random sampling (12.3\% reduction) and MIQP-GL deteriorating from 3624 EUR to 2810 EUR (22.5\% reduction). These perturbed schedules serve as warm-start solutions, enabling the neural penalty predictor to learn linearization strategies that achieve near-optimal performance despite initialization corruption. All DFL methods reported in subsequent tables are trained using the random sampling noise level, representing the most challenging initialization scenario.

The DFL refinement step totals $\approx 1.2$~s in all cases (one LSTM forward pass, $K=7$ QP iterations, and physical simulation). The implementation, however,  differs  in how the warm-start is obtained: (a)~refinement of a piecewise MIQP solution, (b)~real-time scheduling initialized with a fast global linear MIQP, and (c)~MIP-free deployment using a historical schedule retrieved by price-similarity lookup. Scenarios~(a) and~(b) are evaluated below; scenario~(c) is presented in Section~\ref{sec:solver_free}.

Table~\ref{tab:main_results} demonstrates that the DFL framework effectively navigates the trade-off between solution quality and computation time. In scenario~(a), DFL improves ex-post profit by 1.1\% over the MIQP-PW baseline. In scenario~(b), the total wall-clock time is 3.87~s, yielding a 300-fold speedup over MIQP-PW while maintaining profitability within 3.6\% of the piecewise benchmark.

\begin{table}[t!]
\centering
\caption{Performance of DFL Methods, MIQP Benchmarks, and IPOPT Baselines}
\label{tab:main_results}
\small
\begin{tabular}{lcc}
\toprule
\textbf{Method} & \textbf{Ex-post Profit (€)} & \textbf{Time (s)} \\
\midrule
MIQP-GL & $3409 \pm 1433$ & 2.69 \\
\textbf{DFL} (GL-based) & $\textbf{3712} \pm 1509$ & 2.69 + \textbf{1.18} \\
MIQP-PW & $3849 \pm 1545$ & 1205.79 \\
\textbf{DFL} (PW-based) & $\textbf{3890} \pm 1547$ & 1205.79 + \textbf{1.24} \\
\midrule
IPOPT (GL-based) & $3478 \pm 1180$ & 2.69 + \textbf{28.34} \\
IPOPT (PW-based) & $3798 \pm 1498$ & 1205.79 + \textbf{23.76} \\
\bottomrule
\end{tabular}
\end{table}

Fig.~\ref{fig:profit_density_main} shows the density distributions of ex-post profit across scenarios. For GL-based methods (Fig.~\ref{fig:profit_density_main}(a)), DFL exhibits a rightward-shifted distribution in the 3500--4000 EUR range compared to MIQP-GL. For PW-based methods (Fig.~\ref{fig:profit_density_main}(b)), PW-based DFL and MIQP-PW display nearly overlapping distributions with DFL exhibiting slightly higher density near 3900 EUR. The vertical dashed lines confirm the ranking PW-based DFL > MIQP-PW > GL-based DFL > MIQP-GL, where DFL improves solutions within each approximation framework through learned penalty prediction. The close proximity between piecewise methods indicates that DFL achieves near-optimal performance while substantially reducing computational time. As a supplementary comparison, IPOPT baselines initialized with the same mode schedule and warm-start achieve lower ex-post profit than the corresponding DFL variants while requiring 19--24$\times$ longer refinement time, confirming that DFL offers a superior profit--runtime trade-off over a matched nonlinear programming baseline.

Because the system-imbalance and terminal-volume penalties enter the ex-post profit directly, we test whether the conclusions depend on their calibration. We vary the imbalance asymmetry (symmetric $1.0/1.0$, mild $1.5/0.75$, current $2.0/0.5$) and the terminal-volume water value ($0.8\times$, $1.0\times$, $1.2\times$ the median day-ahead price), one axis at a time. For each setting the DFL-PW penalty predictor is retrained (the penalties are its training objective) while MIQP-PW, whose optimization contains no such terms, is re-scored under the same simulator; both methods are thus evaluated identically. As shown in Table~\ref{tab:penalty_sensitivity}, PW-based DFL remains above the MIQP-PW benchmark in every setting, with the margin staying within $1.9$--$2.4\%$. The learned negative SI penalty, which signals that DFL exploits the asymmetric pricing, persists wherever the asymmetry is non-trivial and vanishes, as expected, under symmetric pricing, while a $\pm20\%$ change in the water value leaves the ranking unchanged. The reported advantage is therefore robust to these penalty choices.

\begin{table}[t!]
\centering
\caption{Penalty sensitivity of PW-based DFL vs.\ MIQP-PW.}
\label{tab:penalty_sensitivity}
\footnotesize
\setlength{\tabcolsep}{2.8pt}
\begin{tabular}{lrrrrr}
\toprule
Penalty setting & MIQP-PW & DFL-PW & Gain & DFL SI & DFL Vol \\
 & (€) & (€) & (\%) & (€) & (€) \\
\midrule
Current ($2.0/0.5$, $1.0\times$) & 3793 & 3869 & 1.99 & $-27.8$ & 176.8 \\
SI symmetric ($1.0/1.0$) & 3803 & 3894 & 2.40 & $-51.3$ & 177.5 \\
SI mild ($1.5/0.75$) & 3798 & 3880 & 2.15 & $-38.5$ & 179.6 \\
Water value $0.8\times$ & 3823 & 3902 & 2.08 & $-25.0$ & 152.5 \\
Water value $1.2\times$ & 3763 & 3834 & 1.87 & $-30.1$ & 227.1 \\
\bottomrule
\end{tabular}
\end{table}

\begin{figure}[t!]
    \centering
    \includegraphics[width=\columnwidth]{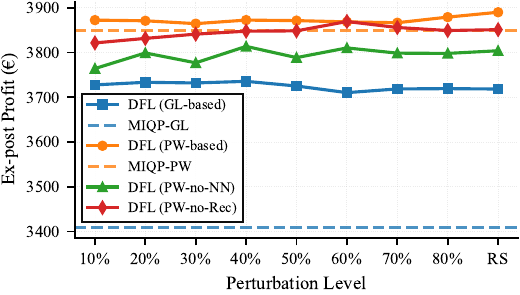}
    \caption{Noise robustness comparison across perturbation levels. DFL methods (solid lines) maintain near-constant performance regardless of initialization quality.  MIQP benchmarks (dashed lines) are shown for reference.}
    \label{fig:noise_robustness}
\end{figure}

Fig.~\ref{fig:noise_robustness} demonstrates that DFL maintains stable ex-post profits across 10\%--80\% perturbation levels (3870--3890 EUR for PW-based; 3710--3735 EUR for GL-based), whereas MIQP baselines degrade by 12.3\%--22.5\%. This confirms that the neural penalty predictor effectively compensates for initialization corruption.

\subsection{Ablation Study: Impact of Neural Network Components}

\begin{figure*}[t!]
    \centering
    \includegraphics[width=\textwidth]{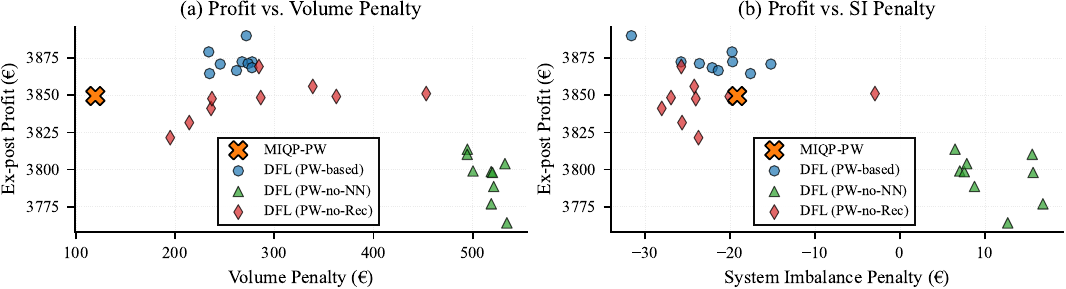}
    \caption{Penalty-profit trade-offs for DFL ablation variants. (a)~Ex-post profit versus volume penalty. (b)~Ex-post profit versus system imbalance penalty. Complete DFL (PW-based) and DFL (PW-no-Rec) achieve superior profit-penalty balance through learned constraint management.}
    \label{fig:penalty_tradeoff_ablation}
\end{figure*}

To isolate contributions of individual architectural components within the DFL (PW-based) framework, two ablated variants are evaluated: DFL (PW-no-Rec) eliminates recursive linearization, executing a single optimization pass, while DFL (PW-no-NN) removes the neural penalty predictor entirely, substituting the best fixed weight triple $(w_p, w_q, w_h)$ identified by exhaustive grid search over $4{,}096$  log-spaced candidates in $[0.01, 100]^3$.

Table~\ref{tab:ablation_results} shows these contributions, where all DFL variants are trained with random sampling initialization. Complete DFL (PW-based) achieves 3890 EUR mean profit in 1.24~s. Removing recursive linearization (DFL-no-Rec: 3851 EUR, 0.46~s) causes 1.0\% degradation, while eliminating the neural network (DFL-no-NN: 3780 EUR, 1.31~s) reduces performance by 2.8\%. The neural network component contributes 22.9\% of the total profit gain over MIQP-GL baseline, while recursive linearization adds an additional 8.1\%, establishing that the neural penalty predictor provides the dominant contribution to noise-robust optimization. Fig.~\ref{fig:noise_robustness} shows this conclusion is consistent across all perturbation levels.

\begin{table}[t!]
\centering
\caption{Ablation Study Results for DFL Components}
\label{tab:ablation_results}
\small
\begin{tabular}{lcc}
\toprule
\textbf{Method} & \textbf{Ex-post Profit (€)} & \textbf{Time (s)} \\
\midrule
DFL (PW-based) & $3890 \pm 1547$ & 1.24 \\
DFL (PW-no-NN) & $3780 \pm 1530$ & 1.31 \\
DFL (PW-no-Rec) & $3851 \pm 1532$ & 0.46 \\
\midrule
MIQP-PW (Benchmark) & $3849 \pm 1545$ & 1205.79 \\
\bottomrule
\end{tabular}
\end{table}

Fig.~\ref{fig:penalty_tradeoff_ablation}(a) reveals the adaptive constraint management enabled by learned penalty prediction. Complete DFL (PW-based) and DFL (PW-no-Rec) accept learned volume penalties averaging 250--280 EUR while achieving ex-post profits exceeding 3860 EUR. This results from the learnable trust region design combined with ex-post profit training, enabling the framework to discover that accepting calculable volume violations during high-revenue periods yields higher net profit. Conversely, DFL (PW-no-NN) incurs substantially higher volume penalties (averaging 515 EUR) and positive SI penalties (5--18 EUR) with lower profits around 3760--3810 EUR, indicating that fixed penalties yield schedules violating both volume targets and power trajectory feasibility during evaluation.

Fig.~\ref{fig:penalty_tradeoff_ablation}(b) demonstrates that DFL learns to exploit asymmetric market pricing, where underproduction incurs penalties at twice the day-ahead price, while overproduction receives compensation at half price. Complete DFL (PW-based) and DFL (PW-no-Rec) achieve negative SI penalty (averaging $-20$ EUR) by biasing schedules toward slight overproduction through learned penalty weights, accepting modest half-price revenue reductions to avoid costly double-price underproduction penalties while maintaining ex-post profits exceeding 3860 EUR. In contrast, DFL (PW-no-NN) incurs positive SI penalties of 5--18 EUR with lower profits around 3760--3810 EUR, indicating that fixed penalties fail to encode this asymmetric cost structure and result in schedules with underproduction. 

These ablation findings establish that learned penalty prediction constitutes the primary mechanism enabling noise-robust optimization and learned profit maximization, while recursive linearization provides incremental value in the profit. For time-critical applications requiring sub-second response, DFL (PW-no-Rec) presents an attractive alternative, retaining the majority of DFL's capabilities through neural penalty prediction alone.

\subsection{MIP-Free Deployment}
\label{sec:solver_free}

The preceding results assume access to an MIQP solver for warm-start generation. In practice, many UPHES operators may lack access to optimization tools or expertise. In deployment scenario~(c), the operator retrieves the most price-similar historical schedule via nearest-neighbor lookup and passes it directly to DFL. We refer to this as \emph{MIP-free} deployment: it eliminates the MIQP that produces the warm-start, while DFL refinement still runs its lightweight convex QP layer (a small ECOS solve per iteration), so the qualifier denotes the absence of the combinatorial solver, not of all optimization.

To validate this workflow, we select 19 held-out days from the remaining 346 days of 2024 Belgian prices by nearest-neighbor price-similarity matching. No retraining is performed: each held-out day is matched to the closest representative date, and that date's pretrained model provides the warm-start mode schedule and performs DFL refinement. This test is strictly distinct from the main results in Section~\ref{subsec:main_results}. There, every representative day is scored by its own model on its own clean reference price profile. Here, every test day is unseen, is routed to the model of a different representative day, and is warm-started from that other day's schedule, so it measures generalization rather than refinement of in-sample data.

\begin{table}[t!]
\centering
\caption{MIP-Free Deployment on 19 Held-Out Price Profiles}
\label{tab:oos_results}
\small
\begin{tabular}{lcc}
\toprule
\textbf{Method} & \textbf{Ex-post Profit (€)} & \textbf{Time (s)} \\
\midrule
MIQP-GL & $2776 \pm 1053$ & 1.37 \\
MIQP-PW & $3249 \pm 1136$ & 1078.6 \\
IPOPT + DB lookup & $2848 \pm 1147$ & 27.61 \\
\textbf{DFL + DB lookup} & $\textbf{2995} \pm 1168$ & \textbf{1.09} \\
\bottomrule
\end{tabular}
\end{table}

\begin{figure*}[t!]
    \centering
    \includegraphics[width=\textwidth]{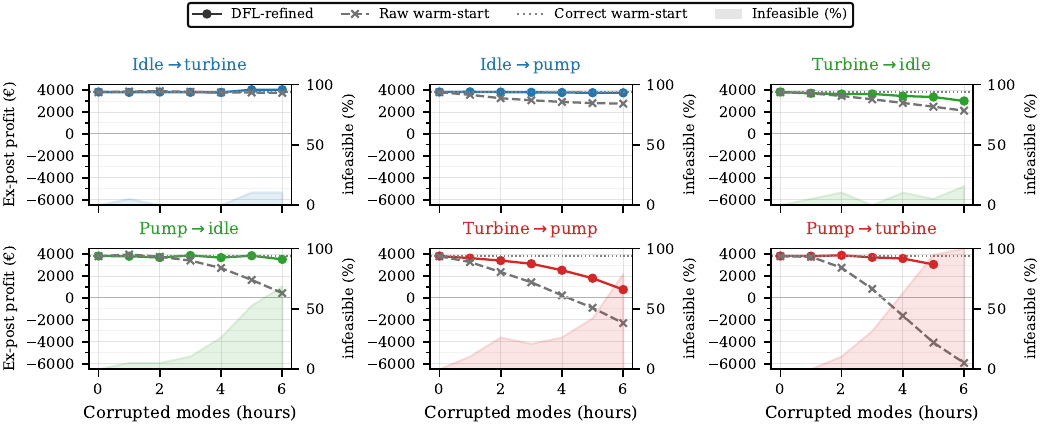}
    \caption{Ex-post profit from six directed warm-start mode errors versus the number of corrupted hours $\rho$. Colored: DFL-refined; gray dashed: raw corrupted warm-start (no refinement); dotted: correct warm-start; shaded (right axis): fraction of the 19 days that become infeasible. Profit is averaged over feasible days.}
    \label{fig:mode_perturbation}
\end{figure*}

Table~\ref{tab:oos_results} shows that DFL + DB lookup retains 92\% of MIQP-PW profit at a 989$\times$ speedup and outperforms MIQP-GL by 7.9\% in mean ex-post profit. Against IPOPT initialized with the same historical warm-start and fixed mode schedule, DFL achieves 5.1\% higher mean profit while being 25$\times$ faster, confirming that the gain stems from the learned penalty predictor rather than the warm-start source.

\subsection{Robustness to Warm-Start Mode Errors}
\label{sec:mode_robustness}

Because the differentiable optimizer locks in the operating mode set by the sign of the warm-start power (Section~\ref{subsec:diff_optimizer}), an incorrect mode cannot be corrected by DFL. Realistic warm-starts rarely disagree on the mode: the fast global-linear MIQP differs from the piecewise reference at only 3.1 hours/day, 88\% of these being idle$\leftrightarrow$active commitments and 12\% turbine$\leftrightarrow$pump sign reversals. To stress-test beyond this regime, we corrupt $\rho$ hours of each clean piecewise reference (baseline profit 3813~EUR) along all six directed transitions (idle$\leftrightarrow$pump, idle$\leftrightarrow$turbine, turbine$\leftrightarrow$pump). Rather than flipping random hours, we flip the ones a planner is most likely to get wrong day-ahead: when wrongly activating an idle hour, the hour where acting is most tempting (highest price for generation, lowest for charging); when wrongly switching off or reversing an active hour, the hour where the unit was barely operating and its correct mode is least obvious. Each curve thus probes the most probable error of its kind rather than an economically self-evident one.

Fig.~\ref{fig:mode_perturbation} reveals a clear severity hierarchy. Spurious commitments are nearly harmless, with DFL holding within 2\% of baseline by driving the wrongly-committed unit to its minimum-power boundary. De-commitments degrade moderately. Sign reversals are most damaging, as the erroneous unit trades against the price signal: the raw warm-start collapses to strongly negative profit (to $-5926$~EUR for pump$\rightarrow$turbine). Wherever a feasible refinement exists, DFL repairs almost all of this loss; for pump$\rightarrow$turbine at $\rho=4$ it recovers from $-1627$~EUR to within 6\% of baseline. The feasible set, however, shrinks rapidly under sign reversals, reaching full infeasibility by $\rho=6$.

This infeasibility is an artifact of the adversarial test design: each error flips the mode in one direction without the compensating counter-switch a genuine re-commitment would entail, so the corrupted schedule no longer conserves the reservoir water balance and the locked-mode QP has no feasible region. The experiment thus delineates a structural requirement rather than a weakness. Given a \emph{feasible} initial mode sequence, DFL recovers near-baseline profit even under the most adverse sign reversals; the only failures are warm-starts that are physically infeasible from the outset. Such a feasible sequence is inexpensive to obtain from any simplified solver (e.g., the global-linear MIQP of Section~\ref{sec:preliminaries}) or a historical operational database (Section~\ref{sec:solver_free}). Removing the fixed-mode dependence altogether motivates the perspective for joint mode-and-trajectory learning in Section~\ref{sec:conclusion}.

\section{Conclusion}
\label{sec:conclusion}

This paper proposed a decision-focused learning framework for UPHES day-ahead scheduling that navigates the trade-off between approximation fidelity and computational speed inherent in mixed-integer nonlinear formulations by integrating neural penalty prediction with recursive local linearization. As a refiner, it improves profit by 1.1\% over piecewise MIQP baselines; as a real-time scheduler initialized with global linear MIQP, it achieves a 300-fold speedup while maintaining profitability within 3.6\% of the piecewise benchmark. An MIP-free deployment mode, where DFL is warm-started from a historical schedule database without any MIQP solver, retains 92\% of MIQP-PW profit at a 989$\times$ speedup on held-out price scenarios.

A key limitation is that the discrete mode commitment is fixed by the warm-start and cannot be corrected by the DFL refiner. Future research directions include end-to-end learning-to-optimize architectures that jointly learn continuous trajectories and discrete mode decisions, eliminating warm-start requirements, and extension to intraday market participation, where shorter decision horizons and sequential clearing mechanisms introduce distinct technical challenges.

\bibliographystyle{IEEEtran}
\bibliography{bibtex/bib/IEEEabrv,reference}

\raggedbottom
\newpage

\begin{IEEEbiography}[{\includegraphics[width=1in,height=1.25in,clip,keepaspectratio]{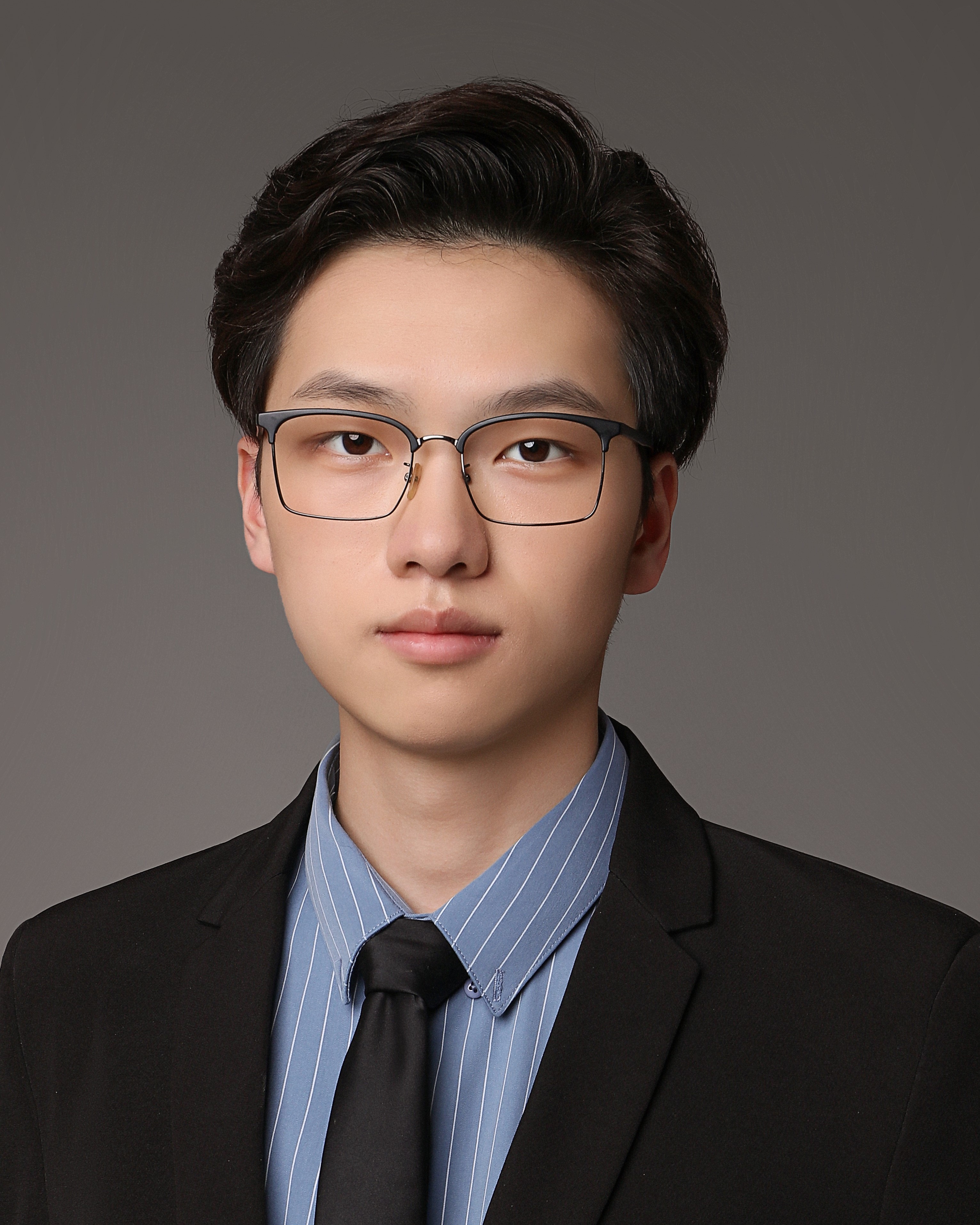}}]{Honghui Zheng}
(Student Member, IEEE) received the B.Eng. degree in water conservancy and hydropower engineering from Hohai University, Nanjing, China, in 2023, and the M.S. degree in systems engineering from Johns Hopkins University, Baltimore, USA, in 2025. He is currently working toward the Ph.D. degree in civil and systems engineering at Johns Hopkins University. His research focuses on the integration of machine learning and mathematical optimization, including learning to optimize, learning to control, and mixed-integer programming, with applications to the scheduling and control of energy systems.
\end{IEEEbiography}

\begin{IEEEbiography}[{\includegraphics[width=1in,height=1.25in,clip,keepaspectratio]{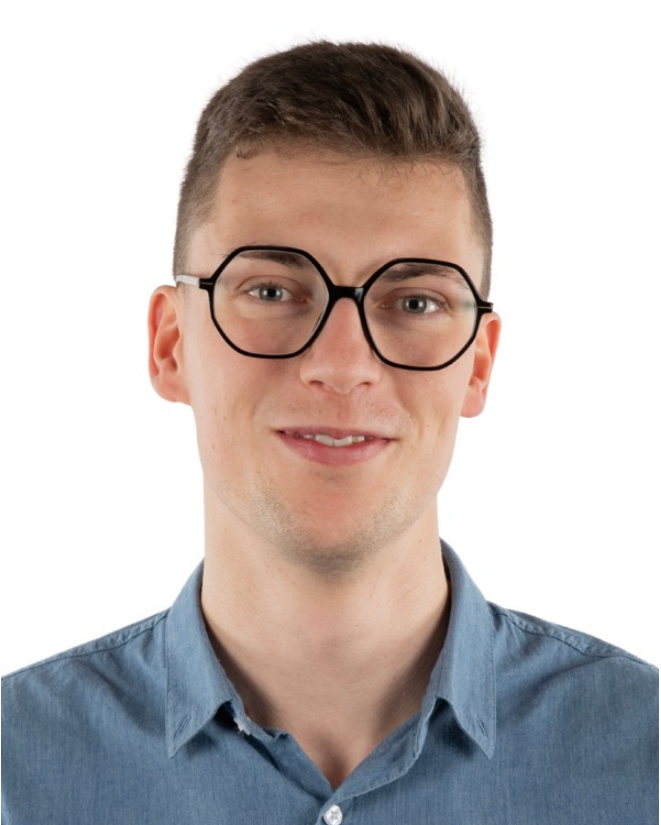}}]{Pietro Favaro}
received his Master’s degree in Electrical Engineering from the University of Mons (UMons), Belgium, in 2022, alongside a Master’s degree in Smart Cities and Communities from Heriot-Watt University, U.K. For his Master’s thesis, he was awarded the Engie Award, the HERA Award in IT, and the KBVE-SRBE Prize. He obtained his Ph.D. in Electrical Engineering from UMons in April 2026, where he is an FNRS–F.R.S. Fellow. In 2023, he was also awarded a one-year fellowship at Johns Hopkins University from the BAEF. His research focuses on the scheduling and control of complex flexibility assets, including pumped hydro energy storage and building energy systems.
\end{IEEEbiography}

\begin{IEEEbiography}[{\includegraphics[width=1in,height=1.25in,clip,keepaspectratio]{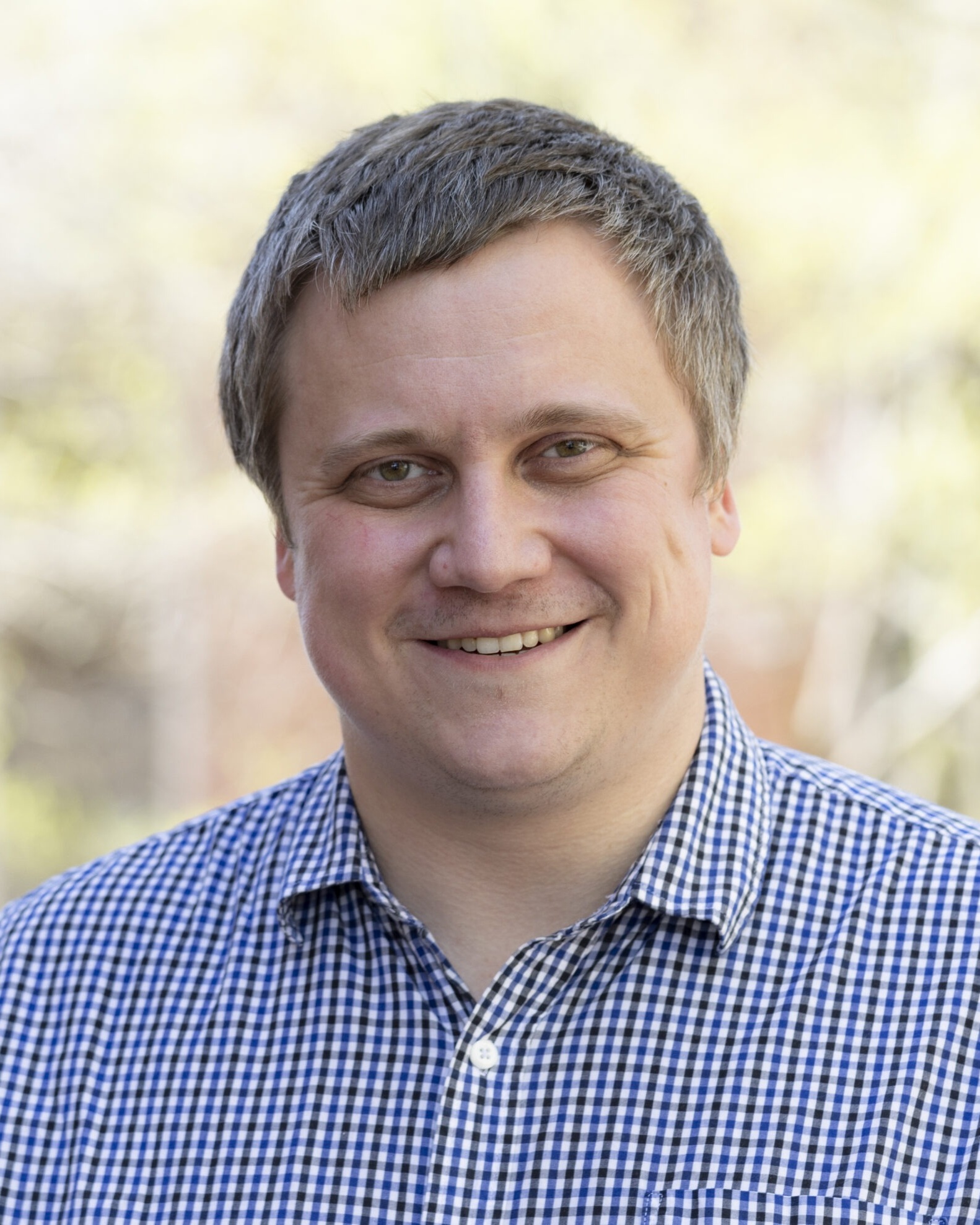}}]{Yury Dvorkin}
(Member, IEEE) received the Ph.D. degree from the University of Washington, Seattle, WA, USA, in 2016. He is currently an Associate Professor with the Departments of Civil \& Systems Engineering and Electrical \& Computer Engineering, Johns Hopkins University, Baltimore, MD, USA. His research interests include modeling, optimization, and machine learning for energy systems planning, operations, and economics. Dvorkin is an Associate Editor for IEEE \textsc{Transactions on Smart Grid} and IEEE \textsc{Transactions on Energy Markets, Policy and Regulation}.
\end{IEEEbiography}

\begin{IEEEbiography}[{\includegraphics[width=1in,height=1.25in,clip,keepaspectratio]{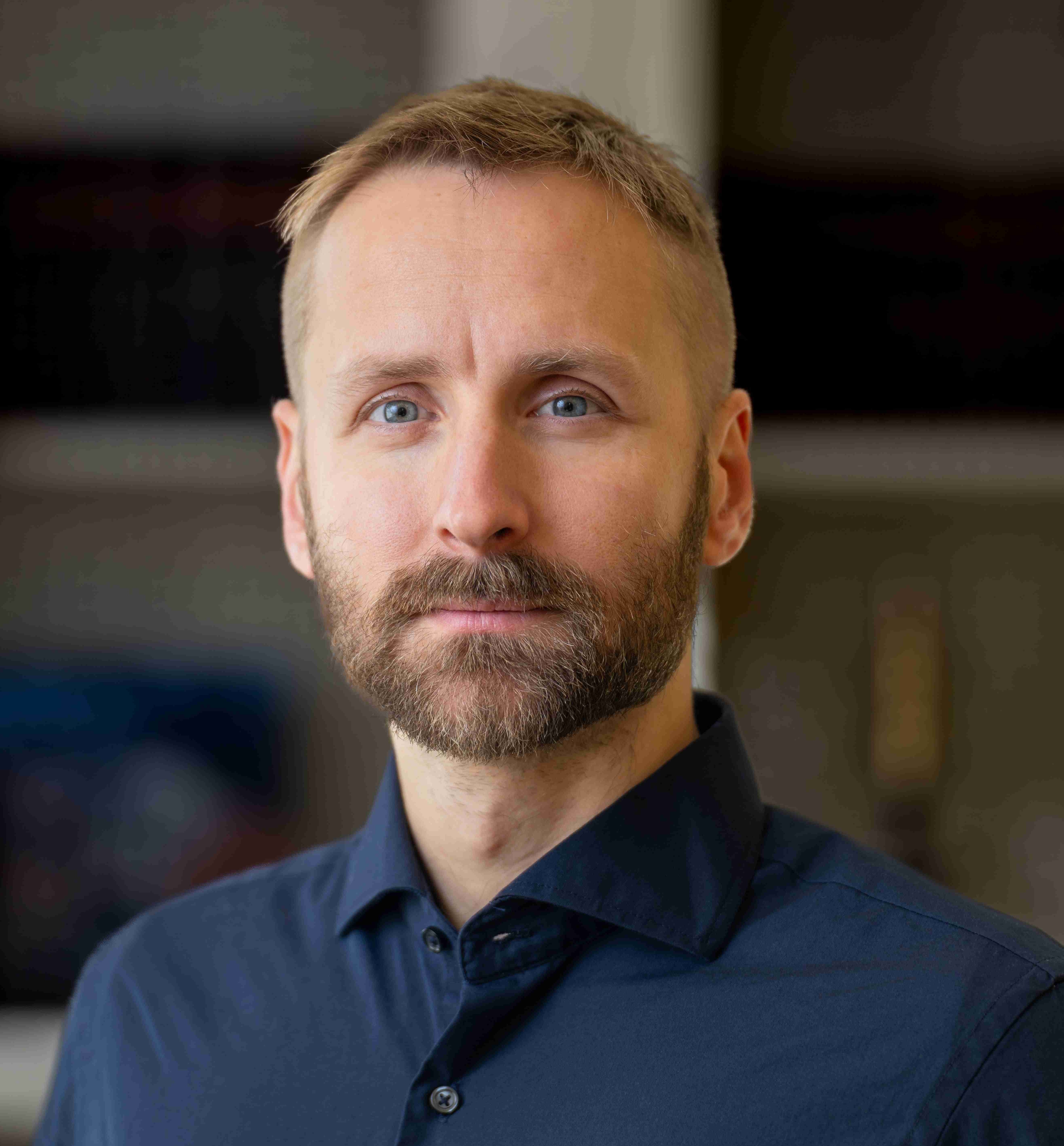}}]{Ján Drgoňa}
(Member, IEEE) is an Associate Professor in the Department of Civil and Systems Engineering, with a secondary appointment in Electrical and Computer Engineering, at Johns Hopkins University. He is a core faculty member of the Ralph S. O'Connor Sustainable Energy Institute (ROSEI) and an associate member of the Data Science and AI Institute (DSAI). Prior to joining Johns Hopkins, he was a Senior Data Scientist at the Pacific Northwest National Laboratory (PNNL), where he continues to hold a joint appointment. He received his Ph.D. in Control Engineering from the Slovak University of Technology and was a postdoctoral researcher at KU Leuven, Belgium. His research focuses on scientific machine learning for modeling, optimization, and control of cyber-physical systems, with applications in sustainable energy.
\end{IEEEbiography}

\end{document}